
\rightline{Imperial/TP/93-94/43}
\noindent
\baselineskip=16pt
\vskip 1cm
\centerline{{\bf Path-integral quantization of $W_\infty$ gravity}}
\centerline{R. Mohayaee}
\centerline{{\it The Blackett Laboratory, Imperial College, London SW7
2BZ, England}}
\vskip 1cm
\centerline{{\bf Abstract}}
\bigskip
\hrule\hfill

\noindent We consider the anomalies of $W_\infty$ gravity in the
context of path-integral
quantization. We derive the ghost-loop anomalies to all orders in
$\hbar$ directly from the
path-integral measure by the Fujikawa method. We also show that in the
matter sector the
higher-loop anomalies can be obtained by implementation of the
Wess-Zumino consistency condition
using the one-loop anomaly. Cancellation of the anomalies between
these two sectors then leaves
the theory anomaly-free.
\vskip 9pt
\hrule\hfill
\bigskip
The $w_\infty$ algebra is a linear higher-spin extension of the
Virasoro algebra which
admits a central term in the spin-2 sector only [1].
The one-scalar realisation of the $w_\infty$ algebra is known as
$w_\infty$ gravity [2].
This realisation satisfies the algebra at the level of double
contractions in the operator-product expansion (but it fails to
generate the algebra at the full quantum level).
 It has been shown [3] that the consequent $w_\infty$ anomalies
 with external matter fields (matter-dependent anomalies) can be
removed by renormalizing the
algebra by the addition of finite local counter terms. This
renormalization changes
 the $w_\infty$ algebra into the $W_\infty$ algebra [3].

 The $W_\infty$ algebra is also a linear
 higher-spin extension of the Virasoro algebra, but unlike $w_\infty$
it admits central terms
for the generators of all integer conformal spins [4]. The $W_\infty$
gravity obtained by the
above renormalization procedure is a one-scalar realisation of the
$W_\infty$ algebra,
  which still may have universal anomalies to all orders in $\hbar$
[4].
These are given by local expressions involving the background gauge
fields only. We shall take
the renormalized $W_\infty$ currents as our starting point for
deriving the universal anomalies.

In our quantization scheme, anomalies
arise from non-invariance of the measure of the path integral [5].
Under symmetries of the action,
the partition function transforms via an ill-defined Jacobian
factor, and this factor, when properly regularized, results in the
anomalies.
The form of the anomalies depends
on the choice of regulator [6] but is restricted by the Wess-Zumino
consistency condition [7]. We show how a certain choice of regulator
 allows us to derive the ghost-loop anomalies to all orders in
$\hbar$. In the matter
sector, only the one-loop anomaly can be derived directly from the
measure. We shall employ the one-loop anomaly in the matter sector
together with
the Wess-Zumino consistency condition to derive all the
higher-loop anomalies.

In order to obtain a proper theory of $W_\infty$ gravity, the
universal anomalies
should also cancel out. In the present case, anomaly cancellation is
controlled by a zeta-function regularization
scheme. The expression for the ghost-sector anomaly involves an
infinite sum to which
different regularization schemes can assign different values. In
previous
works on this subject [8], an appropriate zeta-function regularization
scheme was developed that
ensured the cancellation of diagonal anomalies between the matter and
the ghost sectors. In this
paper, we shall show that off-diagonal anomalies can also be present
in general in the ghost sector,
 prior to the zeta-function regularization. These are not present in
the matter sector and, moreover,
 are not allowed by the Wess-Zumino consistency condition. A proper
zeta-function regularization
 scheme must thus be chosen in such a way as to ensure the
cancellation of diagonal anomalies as
well as the vanishing of all off-diagonal anomalies.

\bigskip
 The action for $W_\infty$ gravity is given by [3]
$$
S={1\over \pi}\int\bigg({1\over
2}\partial\phi\bar\partial\phi-\sum^\infty_{i=0}
A_iV^i\bigg) d^2z \eqno(1)
$$
where $\partial={\partial\over \partial z}=\partial_-$  and
$\bar\partial={\partial\over {\partial\bar z}}=\partial_+$
(adapting Euclidean signature on the world sheet)
and $V^i$ are the renormalized spin $(i+2)$ currents of the form
[3]\footnote\dag{We
set $\hbar$ equal to 1 throughout this paper.}
$$
V^i={(\partial\phi)^{i+2}\over i+2}+{1\over
2}\partial^2\phi(\partial\phi)^i+\cdots\eqno(2)
$$
\noindent  These currents satisfy the operator-product expansion form
of the $W_\infty$ algebra
and generate the following renormalized symmetry transformations of
the $\phi$ field
$$
\eqalign{V^i(z)V^j(w)&=-\sum^\infty_{l=0}f^{ij}_{2l}(\partial_z,\partial_w){
V^{i+j-2l}(w)\over z-w}
-\,4^{-2i}c_i\delta^{ij}\partial_z^{2i+3}{1\over z-w}\, ,\cr
\delta\phi&=\sum^\infty_{l=0}K_l(\partial\phi)^{l+1}-{1\over
2}\partial K_0
+\cdots
\,,\cr}\eqno(3)
$$
where $K_i$ are the parameters of the gauge transformations, $c_i$ are
the
coefficients of the central terms and $f_{2l}^{ij}$ are the structure
constants of the algebra [4]:
$$
f^{ij}_{2l}(m,n)={4^{-2l}\,\phi^{ij}_{2l}\over 2(2l+1)!}
M_{2l}^{ij}(m,n)\,.\eqno(4)
$$
In this expression, $\phi_{2l}^{ij}$ is given in terms of the
saalschutzian hypergeometric
function $\,_4F_3$ :
$$
\phi^{ij}_{2l}=\,_4F_3\left[\matrix{-{1\over 2},{3\over 2},-l-{1\over
2},-l\cr
                \,&\,&;1\cr -i-{1\over 2},-j-{1\over 2},i+j-2l+{5\over
2}\cr}
                  \right],\eqno(5)
$$
and $M$ is a polynomial of degree $2l+1$ in the variables $m$ and $n$
:
$$
M^{ij}_{2l}(m,n)=\sum^{2l+1}_{k=0} {\cal M}^{ij}_{2l,k}\, m^{2l+1-k}\,
n^k\,,\eqno(6)
$$
where
$$
{\cal M}^{ij}_{2l,k}=(-)^k\left(\matrix{2l+1\cr k\cr}\right)
(2i-2l+2)_k[2j+2-k]_{2l+1-k}\,,\eqno(7)
$$
with $(a)_n=\Gamma(a+n)/\Gamma (a)$ and $[a]_n=\Gamma
(a+1)/\Gamma(a-n+1)$.\footnote*{The factors $4^{-2i}$ and $4^{-2l}$
in equations (3) and (4) correct a scaling error in [3].}
The transformation rules of the gauge fields are derived by requiring
the invariance
 of the action and are given by

$$
\delta A_i=\bar\partial K_i+\sum_{j=0}^{i+2l}
\sum_{l=0}^\infty f^{ji-j+2l}_{2l}(\partial _A,\partial _K)A_j
K_{i-j+2l}\,.\eqno(8)
$$
It has been shown [3] that under $\delta\phi$ and $\delta A_i$
transformations, the effective action
is invariant modulo the universal anomalies.

We now perform BRST quantization of this theory. In
the path-integral formalism, one integrates over both matter and gauge
fields.
This necessitates a choice of gauge. The gauge-fixing procedure
is conveniently carried out using a background-field gauge
$A_i=A_i^{back}$
which is imposed by a Lagrange multiplier $\pi_i$ in the action. After

gauge fixing, the action, symmetries and partition function are
replaced  by their
BRST counterparts
$$
\eqalign{
S_{BRST}&= S-{1\over \pi}\int\sum_{i=0}^{\infty}\bigg( b_i
\bar\partial c_i +
\sum_{l=0}^\infty\sum_{j=0}^{i+2l}b_if^{ji-j+2l}_{2l}
(\partial_A,\partial_c)A_jc_{i-j+2l}-
\pi_i(A_i-A_i^{back})\,\bigg),\cr
\delta\phi&=\sum_{l=0}^\infty c_l(\partial\phi)^{l+1}-{1\over 2}
\partial c_0 +\cdots\,,\cr
\delta A_i&=\sum_{j=0}^{i+2l}
\sum_{l=0}^\infty\bigg(\bar\partial c_i+f_{2l}^{ji-j+2l}(\partial_A,
\partial_c)A_j c_{i-j+2l}\,\bigg),\cr
\delta c_i&=-\sum_{j=0}^{i+2l}
\sum_{l=0}^\infty{1\over
2}f_{2l}^{ji-j+2l}(\partial_{c_j},\partial_{c_{i-j+2l}})
c_j c_{i-j+2l}\,,\cr
\delta b_i&=\pi_i\,,\cr
\delta\pi_i&=0 \,,\cr
Z&=\int{\cal D}\phi{\cal D}A{\cal D}b{\cal
D}c\,\exp\,(S_{BRST})\,.\cr}
\eqno(9)$$
The BRST symmetries of $c_i$ and $b_i$ follow from nilpotence of
$\delta A_i$
and invariance of the action, respectively.

Next we show how the non-invariance of the path-integral measure under
the symmetries
of the action gives rise to anomalies and how these anomalies can be
calculated.
 Anomalies (${\cal A}$) originate from the breakdown of BRST
invariance in the path-integral
measure. These appear in the Ward identities as a logarithm of the
Jacobian $({\cal J})$ of
the BRST transformation :
$$
{\cal A}=\ln{\cal J}=\ln\,{\rm det}\sum_{n,k=0}^\infty
{\partial^l(\phi+\delta\phi,A_n+\delta A_n,c_n+\delta c_n,b_n+\delta
b_n)\over
 \partial(\phi,A_k,c_k,b_k)}\,,
\eqno(10)$$
where $\partial^l$ denotes left differentiation and the explicit form
of the anomaly
 ${\cal A}$ is
$$
{\cal A}={\rm tr}\,\sum_{n,l=0}^\infty\sum_{j=0}^{i+2l}
\left(\matrix{{\cal N}_{\phi\phi}&0&{\cal N}_{\phi c}\cr
0&{\cal N}_{gg}&
{\cal N}_{gc}
\cr
0&0&{\cal N}_{cc}\cr}\right)
\delta^2(z-w).\eqno(11)
$$
where
$$
\eqalign{{\cal
N}_{\phi\phi}&=(l+1)c_l(\partial\phi)^l\partial+\cdots,\cr
{\cal N}_{cc}&=-{\cal N}_{gg}=f_{2l}^{j
n-j+2l}(\partial_c,\partial_\delta)c_j,\cr}\eqno(12)
$$
and the expressions for ${\cal N}_{\phi c}$ and ${\cal N}_{gc}$ follow
in exactly the same way.
Since ${\cal N}_{\phi\phi}$  cannot be written down explicitly
as an infinite sum, we shall initially consider only terms with no
factor of $\hbar$ in the matter sector.
The trace operation in (11) includes integrations over $z$ and $w$.
Because it involves the product of a delta function and its
derivative, the anomaly ${\cal A}$ is an
ill-defined object that requires  regularization.

In our regularization scheme, the delta function in (11) is expanded
in terms of the complete set of
 eigenstates $\varphi_n$ of a hermitian operator $H$ as follows
$$
\delta^2(z-w)=\lim_{M\to\infty}\sum^{\infty}_{-\infty}\varphi^*_n(w)\,e^{-H/M^2}\,\varphi_n(z)\,.

\eqno(13)
$$
This scheme in effect replaces the $\delta$ function in (11) by a
regulator:
$$
{\cal A}=\lim_{M\to\infty}{\rm tr}\sum_{n,l=0}^\infty\sum_{j=0}^{i+2l}
\bigg({\cal N} \,e^{-H/M^2}\bigg),\eqno(14)
$$
where ${\cal N}$ stands for the matrix in (11).

So far, there is no restriction on the choice of the Hermitian
operator $H$; however,
our final answer for ${\cal A}$ will have to satisfy the Wess-Zumino
consistency
condition [7]. It has been shown [6] that a set of operators
appropriate for use as Fujikawa
regulators arises in the Pauli-Villar regularization scheme and that
these
lead to anomalies satisfying the consistency condition [9].
The Fujikawa regulators found in this way are constructed from the
propagators of the theory.
 Non-propagating fields like the gauge fields are regularized
  using operators acting on antighosts in the action; the latter
transform into Lagrange
multipliers and thus do not contribute to the anomalies.
In the Fujikawa scheme one obtains the operator $H_\phi$ acting on the
field $\phi$ from $\partial^2
S_0/\partial\phi^2$ :
$$
H_{\phi}=-\partial\bar\partial+\sum^{\infty}_{l\geq0}\partial\big[(l+1)

A_l(\partial\phi)^l\partial\big]\,,\eqno(15)
$$
where $S_0$ is the part of $S$ which is of zeroth order in $\hbar$.
Since Fujikawa regularization is restricted to the one-loop universal
anomalies,
the Jacobian and the regulators in the matter sector can initially be
limited to contain terms of zeroth order in $\hbar$ only.
 The operator $H_\phi$, containing a d'Alembertian, is Hermitian
and positive semi-definite and is acceptable as it stands [6].
On the other hand, the ghost sector operator derived from
${\partial^2S_{BRST}\over \partial b_k\partial c_n}$, given by

$$
D=\left(\matrix{0&D_1\cr D_1^\dagger&0\cr}\right)\eqno(16)
$$
with $D_1=-\bar\partial-\sum_{r=0}^\infty f_{2r}^{2(l+r)-j
n}(\partial_A,\partial) A_{2(r+l)-j}$,
is not positive semi-definite. Consequently, one needs to use the
square of $D$
 as the regulator in the ghost/anti-ghost sector.\footnote*{Hermitian
conjugation and the inner product are defined by
 $(H\psi,\varphi)=(\psi,H^\dagger\varphi) $ and
$(\psi,\varphi)=\int\bar\psi\varphi$.}
Remembering that our present concern is with a chiral gauging of
$W_\infty$, we need from the
regulators only the terms involving $A$ and not $\bar A$. Making the
corresponding truncation
in $D^2$, one obtains the following expression for the Hermitian
operator $H$ :
$$
H=\left(\matrix{H_\phi&0&0\cr
           0&D_1D_1^\dagger&0\cr  0&0&D_1^\dagger
D_1\cr}\right)_{|_{chiral}}\,.\eqno(17)
$$
Because the regulator is diagonal, the calculation is simplified and
the anomaly ${\cal A}$, given in (11), dissociates into its individual
components in the matter, ghost and gauge sectors :
$$
{\cal A}={\cal A}_\phi+{\cal A}_c+{\cal A}_g\,,
$$
in which
$$
  \eqalign{{\cal A}_\phi&=
                         \lim_{M\to\infty}\sum_{l=0}^{\infty}
                          {\rm tr}\bigg({\cal N}_{\phi\phi}
                          e^{-H_\phi /M^2}\bigg),\cr
            {\cal A}_c&=
                         \lim_{M\to\infty}\sum_{l,n=0}^\infty\sum_{j=0}^{n+2l}
                         {\rm tr}\bigg({\cal N}_{cc} e^{- D_1^\dagger
D_1/M^2}\bigg),\cr
            {\cal A}_g&=
                          \lim_{M\to\infty} \sum_{l,n=0}^\infty
                         \sum_{j=0}^{n+2l}{\rm tr}\bigg({\cal N}_{gg}
                          e^{-D_1D_1^\dagger/M^2}\bigg).\cr}\eqno(18)
$$

 One can now use the completeness relation for the basis $\varphi_n$
given
in (13) to evaluate the above traces in a plane-wave basis. After a
relatively long
 calculation we obtain
$$
{\cal A}_\phi=\sum^\infty_{l\geq 0}\sum^\infty_{j\geq
0}\int{(l+1)(j+1)\over 12\pi}
c_l(\partial\phi)^l\partial^3(A_j(\partial\phi)^j)\,.\eqno(19)
$$
All the $\phi$-dependent terms in the above expression cancel against
the variation of
the $W_\infty$ action [3] leaving the one-loop universal anomaly
$$
{\cal A}_\phi^u={-1\over 6\pi}\int\,dz\, c_0\partial^3 A_0\,.\eqno(20)
$$

\noindent Similar calculations yield ${\cal A}_c$ and ${\cal A}_g$.
The relevant quantity,
${\cal A}_c+{\cal A}_g$, is given by
$$
\eqalign{{\cal A}_c+{\cal A}_g &
={-1\over \pi}\sum_{l,r,n=0}^\infty\sum_{j=0}^{2l+n}\sum_{s=0}^{2l+1}
\sum_{s\prime=0}^{2r+1}\sum_{m=s\prime}^{s+s\prime}\cr
\,&\bigg({(-1)^{m+s} s!
\over (m+1)(s+s\prime-m)!(m-s\prime )!}
\Phi^{j n-j+2l}_{2l,s}\Phi^{2(r+l)-j n}_{2r,s\prime}\bigg)
c_j\partial^{2(r+l)+3}A_{2(r+l)-j},\cr}\eqno(21)
$$
where the $\Phi$'s are the deformed structure constants of the
$W_\infty$ algebra :
$$
\Phi^{\beta\gamma}_{2\alpha,\lambda}=\,\phi^{\beta\gamma}_{2\alpha}\,
{\cal M}^{\alpha\beta}_{2\alpha,\lambda}\,.\eqno(22)
$$

\noindent In the above expression  $\phi_{2\alpha}^{\beta\gamma}$ is
the hypergeometric function $_4F_3$ and ${\cal M}$ is
 the coefficient of the polynomial part of the structure constant.
These are given in (5) and (7)
respectively.
The expression for ${\cal A}_c+{\cal A}_g$ contains terms of all
orders in  $\hbar$.\footnote*{
There is a factor $\hbar^{l+r+1}$ in (21) which we have set equal to
1.}
Since the ghost action is entirely determined by the structure
constants of the algebra, it is bilinear in
ghosts.
So, although anomalies will occur at all orders in $\hbar$, they do
not
arise from multi-loop diagrams. This explains why these anomalies can
be derived directly from the measure of the path integral.

For the spin-2 ghost sector anomaly, we obtain
$$
({\cal A}_c+{\cal A}_g)_{(2)}={1\over 12\pi}\int \sum^\infty_{n= 0}
[6(n+1)^2 +6(n+1)+1]c_0\partial^3A_0\,.\eqno(23)
$$
The above summation is strongly divergent. In order to regularize the
sum, a zeta-function
regularization scheme was used in [8].

The generalized Riemann zeta function is defined by
$$
\zeta(s-l,\alpha)\equiv\sum^\infty_{n=0}(n+\alpha)^{l-s} \ \ \ \ \ \ \
\ \ \ \alpha\neq 0, -1, -2\cdots\ \ \ \ \ s>l+1 \eqno(24)
$$
and is calculated using Bernoulli's polynomials [12]. The formally
divergent sum $\sum^\infty_{n=0}
(n+\alpha)^l$ can be interpreted as the analytic continuation of the
zeta function to $s=0$
 $\big(\zeta(-l,\alpha)\big)$. The regularized value of the divergent
sum depends on
 the parameter $\alpha$. Because the sum in (23) is not absolutely
convergent, there is
an ambiguity in the grouping of terms before regularization. This
leads to regularization schemes employing
different $\alpha$'s and consequently resulting in different values
for the divergent sum.
 This ambiguity is resolved
in [8] for the spin-2 anomaly by requiring that the result obtained by
regularizing the sum in (23)
respects the symmetry under the interchange $n+1\rightarrow-(n+2)$
existing in the
$(b,c)$ system. Upon choosing a scheme of regularization that respects
this symmetry one
obtains the value 2 for the infinite sum in (23). This is the value of
the ghost central charge.

As one goes on to higher spins, the freedom in choosing a scheme of
regularization increases and the
ambiguities multiply. However, the regularization schemes for
different spins must be consistent with one
another since, as we will show later, all the higher-spin anomalies
can be obtained from the
spin-2 anomaly by the Wess-Zumino consistency condition. An extension
of the spin-2 regularization
scheme was proposed in [13] where it was shown to give consistent
results at least up to the
spin-18 level for the diagonal anomalies (anomalies of the form
$c_i\partial^{2i+3}A_i$).

One also needs, however, to consider off-diagonal
anomalies (anomalies of the form $c_j\partial^{2(r+l)+3} A_{2(r+l)-j}$
for ${r+l}\neq j$).
Although overlooked in the first reference of [8], these anomalies, as
can be seen from (21),
do generically exist in the theory.\footnote*{Off-diagonal anomalies
can also be derived
 from the BRST charge $(Q)$. The operator $Q^2$ contains off-diagonal
anomalies of the form
$c_j\partial^{2(l+r)+3}c_{2(r+l)-j}$ for $r+l\neq j$.}
Since the anomalies in the matter sector are purely
diagonal, the presence of off-diagonal ghost-sector anomalies is
problematic.
We shall show, however, that these off-diagonal anomalies are not
allowed by the
 Wess-Zumino consistency
condition. This provides a strong reason for choosing a certain scheme
of zeta-function
regularization in preference to others.

In the preceding calculation, we derived the one-loop matter sector
anomaly  $({\cal A})$
from the measure. We will now show that all of the higher matter-loop
anomalies can be derived by implementation of the Wess-Zumino
consistency condition on ${\cal A}_\phi^u$.

 We start with the commutation relation
$$
[\delta_{c_m},\delta_{c_n}]A_i=\delta_{c_{[m,n]}} A_i+ g_{m,n,j}A_j\ \
\ \ \ \ \ j\neq i\,,\eqno(25)
$$
where $c_{[m,n]}=(n+1)c_n\partial c_m-(m+1)c_m\partial c_n.$
This relation is then applied to the generator of
one-particle-irreducible diagrams with
external gauge fields only ($\Gamma$) :
$$[\delta_{c_m},\delta_{c_n}]\Gamma=\delta_{c_{[m,n]}}\Gamma +\int
g_{m,n,j}A_j
{\delta\Gamma\over \delta A_i}\,.\eqno(26)$$

This form of the Wess-Zumino consistency condition can be used for
diagrams with
external gauge-field lines only. In the BRST formalism, the ghosts
play the roles of the parameters of
the gauge transformations. As a result, anomalies which were
originally functionals of the gauge
fields only now become functionals of the ghost fields as well. In the
case of $W_\infty$ gravity where
the gauge-field transformations are linear, the BRST
transformations are obtained directly from the gauge transformations
but with a substitution
of ghosts for the original parameters. Consequently, the original
Wess-Zumino commutator form of
the consistency condition
 can be used, modulo the treatment of the ghosts as gauge parameters.
It should be noted
that this simplified treatment of the consistency conditions is not
always possible. In more complicated theories
like $W_3$ gravity, the BRST transformations cannot simply be
rewritten as gauge transformations with ghosts replacing the original
parameters. In [10] an anti-bracket form of the Wess-Zumino
consistency condition was developed
from the Ward identities and used for analysing the anomalies of $W_3$
gravity. In that case, the
more elaborate treatment of the consistency conditions is necessary
because the gauge fields
transform nonlinearly into the scalar fields. In
$W_\infty$ gravity, on the other hand, the BRST transformations of the
gauge fields
involve terms at most linear in the gauge and ghost fields. As a
result, the
anti-bracket form of the Wess-Zumino consistency condition given in
[10]
reduces to the commutator form (25) of the condition, by virtue of the
linearity of the $W_\infty$ algebra.

 Let us now proceed with our calculations, keeping in mind the
limitations
imposed by the above form of the Wess-Zumino consistency conditions.
The second term in (26) is a non-local expression. All the non-local
anomalies should
cancel independently of the local ones. For this reason, we shall not
consider them in our
search for the local universal anomalies.
 Insertion of the spin-2 universal anomaly ${\cal A}_\phi^u$ (20) into
(26) gives
$$
{1\over
6\pi}\int\partial^3c_0f^{2l-m\,m}_{2l}(\partial_A,\partial_c)A_{2l-m}c_m=\int

c_m(\delta_{c_0}-(m+2)\partial c_0-c_0\partial)\Lambda_{c_m}\,,
\eqno(27)$$
where we have used the notation
$$\delta_\alpha\Gamma={\cal
A}_\alpha=\int\alpha\Lambda_\alpha\,.\eqno(28)$$

\noindent On the basis of dimensional analysis we can see that
$\Lambda_{c_m}$ must be of the form

$$
\Lambda_{c_m}=\alpha_{l,m}\partial^{2l+3}A_{2l-m}\,,\eqno(29)
$$

\noindent where $\alpha_{l,m}$ is a coefficient to be found.
 To derive the form of $\alpha_{l,m}$ this expression must be inserted
into the consistency condition
(27).
The end result is that there is no consistent value for $\alpha$ if
$l\neq m$. That is
to say, off-diagonal anomalies are not allowed by the consistency
condition.
Since the consistency condition has exactly the
same form in the ghost sector; off-diagonal anomalies are not allowed
in the
ghost sector either. As a result a proper zeta-function regularization
is one that not only
 gives the right value for the central charge but also
puts the coefficents of all off-diagonal anomalies in the ghost sector
equal to zero.
It is not easy to find a scheme that works for all spins. As explained
earlier,
the regularization ambiguities multiply as we go to higher orders and
a generic zeta-function
regularization will not lead to cancellation of all off-diagonal
anomalies. However, appropriately
chosen schemes that give cancellation of the off-diagonal anomalies do
exist; an example of such
a scheme is the one given in [13]. Thus, our Wess-Zumino consistency
condition considerations provide
a stringent requirement on the summation scheme used to combine the
contributions from
different spin sectors to the anomalies. This consistency condition is
important because the sum of the contributions from the
different spin sectors is otherwise ambiguous. This ambiguity is more
serious than that encountered in the
use of zeta functions to regulate Feynmen diagrams, because there one
has the restrictive requirement
of uniformly regulating different divergences and also preserving as
much as possible the symmetries
of the action [14]. The Wess-Zumino consistency condition appears to
play an analogous role in
restricting the summation of the contributions to the anomalies.

The diagonal part of anomalies has the following coefficients
$$\alpha_{m,m}={-\Phi^{m m}_{2m,0}\over 12\pi(2m+1)!(m+1)}\, ,
\eqno(30)$$
where $\Phi^{m m}_{2m,0}$ is given in (22).
\noindent Because of the simple relationship between the indices on
$\Phi$, the hypergeometric function $_4F_3$ appearing in the
definition of
  $\Phi^{m m}_{2m,0}$  is reducible to the hypergeometric function
$_2F_1$ :
$$
_4F_3\left[\matrix{-{1\over 2}&{3\over 2}&-m\cr \,&\,&\,&\,&;1\cr
-m-{1\over 2}&-m-{1\over 2}&{5\over 2}\cr}\right]={(3)_m\over
 ({5\over 2})_m}\,_2F_1\left[\matrix{-{1\over 2}&-m\cr
\,&\,&\,;1\cr-m-{1\over 2}\cr}\right] .\eqno(31)
$$
The hypergeometric function $_2F_1$ , on the other hand, can be
calculated using Gauss'
 law for hypergeometric functions [11].
Paying special attention to hypergeometric functions with negative
elements,
 we obtain the final answer for $\alpha_{m,m}$ :
$$
\alpha_{m,m}=-2^{2m-2}{m!(m+2)!\over (2m+1)!!(2m+3)!!}\ .\eqno(32)
$$
Consequently the final local matter anomalies to all loop orders are
given by:
$$
{\cal A}_\phi^{\rm
u}=\int\sum_m\alpha_{m,m}c_m\partial^{2m+3}A_m\,.\eqno(33)
$$
 We can check order-by-order that ${\cal A}_\phi^{\rm u}$ cancels
against the regularized anomalies in the ghost sector
$({\cal A}_c +{\cal A}_g )$.

We therefore conclude that the Fujikawa method taken together with
the Wess-Zumino consistency condition provides a complete program of
anomaly derivation to all
orders in $\hbar$ in $W_\infty$ gravity. Using the consistency
condition and
the one-loop result for matter and ghost fields, we have demonstrated
the order-by-order cancellation
of anomalies for the single-scalar realisation of $W_\infty$ gravity.
To achieve
this cancellation, one needs to sum up the ghost contributions to the
anomalies using a zeta-function regularization scheme that is selected
so as to
ensure the validity of the consistency condition.
\bigskip
\noindent{\bf Acknowledgement}
\bigskip
I am grateful to Kelly Stelle for suggesting the
problem, supervising me in the course of its progress and making
numerous detailed comments on the manuscript. I would also like to
thank Arley Anderson, Luis Garay and
Bill Spence for helpful discussions.
\bigskip

\frenchspacing
\noindent{\bf References}
\bigskip
\item{[1]} I. Bakas, Phys. Lett. {\bf B228} (1989) 57.
\item{[2]} E. Bergshoeff, C.N. Pope, L.J. Romans, E. Sezgin, X. Shen
      and  K.S. Stelle,
       Phys. Lett. {\bf B243} (1990) 350.
\item{[3]} E. Bergshoeff, P.S. Howe, C.N. Pope, E. Sezgin, X. Shen and
K.S. Stelle, Nucl. Phys.
{\bf B363} (1991) 163.
\item{[4]} C.N. Pope, L.J. Romans and X. Shen, Phys. Lett. {\bf B236}
(1990) 173;
\item{   } C.N. Pope, L.J. Romans and X. Shen, Nucl. Phys. {\bf B339}
(1990) 191.
\item{[5]} K. Fujikawa, Phys. Rev. Lett. {\bf 42}, (1979) 1195; Phys.
Rev. Lett. {\bf 44}, (1980) 1733;
     Nucl. Phys. {\bf B226}, (1983) 437.
\item{[6]} A. Diaz, W. Troost, P. Van Nieuwenhuizen and A. Van
Proeyen,
     Int. J. Mod. Phys. {\bf A4} (1989) 3959;
  \item{ } W. Troost, P. Van Nieuwenhuizen and A. Van Proeyen, Nucl.
Phys. {\bf B333}
     (1990) 727;
  \item{ }M. Hatsuda,  W. Troost, P. Van Nieuwenhuizen  and A. Van
Proeyen, Nucl. Phys.
     {\bf B335} (1990) 166.
\item{[7]} J. Wess and B. Zumino, Phys. Lett. {\bf B37} (1971) 95;
 \item{ } W. Bardeen  and B. Zumino, Nucl. Phys. {\bf B244} (1984)
421.
\item{[8]} K. Yamagishi, Phys. Lett. {\bf B266} (1991) 370;
\item{ }   C.N. Pope, L.J. Romans and X. Shen, Phys. Lett.{\bf B254}
(1991) 401.
\item{[9]} S. Vandoren and A. Van Proeyen, Nucl. Phys. {\bf B411}
(1994) 257.
\item{[10]} R. Mohayee, C.N. Pope, K.S. Stelle and K.W. Xu, Preprint,
"Canonical BRST quantisation of
   worldsheet gravities", CTP TAMU-18/94, Imperial/TP/93-94/31,
hep-th/9404170.
\item{[11]} L.J. Slater, {\it Generalized hypergeometric functions}
(Cambridge Univ. Press, Cambridge, 1966).
\item{[12]}I.S. Gradshteyn and I.M. Ryzhik, {\it Tables of integrals,
series and products}
(Academic Press, San Diego, 1980).
\item{[13]} C.N. Pope, L.J. Romans, E. Sezgin and X. Shen, Phys.
Lett.{\bf B256} (1991) 191;
\item{ }    C.N. Pope, L.J. Romans and X. Shen, in: {\it Strings 90},
eds R. Arnowitt, K. Bryan, M.J. Duff, D. Nanopoulos,
         C.N. Pope, E.Sezgin, (World Scientific, Singapore, 1991);
\item{ } C.N.Pope, Trieste summer school lecture notes, 1991.
\item{[14]} P. Ramond, {\it Field theory: A modern primer},
(Addison-Wesley, Reading, MA, 1990).

\end